\documentstyle[12pt,aps,epsf,amsfonts,amssymb]{revtex}
\def\mycap#1{ 
\parbox[h]{\textwidth}{\vskip 0.4cm
\footnotesize \baselineskip 0.mm  #1 } }

\begin{document}
\author{{\bf Miodrag L. Kuli\'{c}} and {\bf Martin Endres }}
\address{Physikalisches Institut, Theorie III, \\
Universit\"{a}t Bayreuth, 95440 Bayreuth, Germany  }
\title{Ferromagnetic Semiconductor - Singlet (or Triplet) Superconductor - Ferromagnetic 
Semiconductor Systems as 
Possible Logic Circuits and Switches}
\date{07 December 1999}
\maketitle

\begin{abstract}
We consider thin superconducting ($S$) films of thickness $d$ $\ll \xi _{0}$,
sandwiched between two ferromagnetic semiconducting insulators ($FI$)
with differently orientated
magnetizations - the $FI-S-FI$ system. 
We calculate the dependence of the superconducting critical
temperature on the orientation of the magnetization in the
insulators and on the thickness of the superconducting film. The
calculations are done for singlet as well as triplet superconductors. In the
singlet case $T_{c}$ depends on the relative orientation of the left and right
magnetization only, while in the triplet case $T_{c}$ depends on the absolute 
orientation of magnetization. The latter property can serve as a kind of
spin-spectroscopy of triplet and unconventional superconductors, for instance 
in resolving the structure of the triplet order parameter in the recently
discovered layered superconductor $Sr_{2}RuO_{4}$.
The possibility of logic circuits and switches, which are based on the 
$FI-S-FI$ systems with arbitrary orientation of magnetizations
in $FI$ films, is analyzed too.
\end{abstract}

\newpage

\section{\ Introduction}

The physics of magnetic superconductors is especially interesting because of
the competition between magnetic order and superconductivity. Important
progress in the field started with the discovery of the magnetic
superconductors $RERh_{4}B_{4}$, $REMo_{6}S_{8}$, $REMo_{6}Se_{8}$, etc.
where the rare earth ions, $RE$, are regularly distributed in the crystal
lattice. After intensive experimental \cite{Maple} and theoretical \cite
{BuBuKuPaAdvances} investigation it turned out that in many of these systems
superconductivity (with the critical temperature $T_{c}$) coexists with
antiferromagnetic order (with the critical temperature $T_{N}$). One usually
has $T_{N}<T_{c}$. However, due to their antagonistic characters singlet
superconductivity and simple ferromagnetic order can not coexist in bulk
samples with realistic physical parameters as it was shown theoretically 
\cite{BuBuKuPaAdvances}, \cite{BuBuKuPa}, \cite{BuKuRu}. In fact, under
certain physical conditions ferromagnetic order is transformed into a spiral
or domain-like structure in the presence of superconductivity, as it was
observed in $ErRh_{4}B_{4}$, $HoMo_{6}S_{8}$, $HoMo_{6}Se_{8}$ \cite{Maple}, 
\cite{BuBuKuPaAdvances}. At the same time the superconducting order
parameter is suppressed even in the presence of the modified ferromagnetism
in such a way that they may coexist in some limited temperature interval 
(in $ErRh_{4}B_{4}$, $HoMo_{6}S_{8}$) or even down to $T=0K$
(in $HoMo_{6}Se_{8}$) depending on system parameters. For more details see
Refs. \cite{BuBuKuPaAdvances}, \cite{BuBuKuPa}, \cite{BuKuRu}. Recently, the
coexistence of superconductivity and nuclear magnetism was found
experimentally in $AuIn_{2}$ \cite{Pobell} and explained theoretically \cite
{KuBuBu} in terms of a spiral or a domain-like structure . There is also
evidence for the coexistence of ferromagnetism, which appears at $T_{M}=137$ 
$K$, and superconductivity, which sets in at $T_{c}=45$ $K$, in the layered
perovskite superconductor RuSr$_{2}$GdCu$_{2}$O$_{8}$ \cite{Braun}, \cite
{Bernhard}. The specificity of this material is that the ferromagnetic ($F$)
order is present in Ru-O planes while superconducting ($S$) pairing
dominates in Cu-O planes.

Recently, various artificial multilayer $SF$\ systems were prepared in the
form of bilayer or trilayer systems or superlattices consisting of
ferromagnetic and superconducting films deposited on each other. In all of
these artificial structures, for instance in $Nb/Gd$ \cite{Jiang1}, \cite
{Chien} and $Nb/CuMn$ multilayers \cite{Mercaldo}, in $Fe/Nb/Fe$ trilayers 
\cite{Zabel1}, \cite{Zabel2}, in $Nb/Gd/Nb$ trilayers \cite{Jiang2},
superconductivity and ferromagnetism are spatially separated. In spite of
this fact $T_{c}$ is suppressed  in these systems due to the combined 
paramagnetic and proximity effect in the normal metal.
The theoretically predicted oscillation of $T_{c}$ \cite{Buzdin} as a function
of the thickness of the ferromagnetic film in $SFS$ multilayers was observed 
in $V/Fe$ \cite{Wong} and in $Nb/Gd$ \cite{Jiang1}, \cite
{Strunk} superlattices, while in some artificial structures of $V/Fe$ \cite
{Koor} and $Nb/Fe$ \cite{Verb} this effect was absent. The oscillation of $%
T_{c}$ is due to the $\pi $-Josephson junction in the $SFS$ system \cite
{Buzdin}, \cite{Bul}. Recently, several theoretical papers were published
which study transport properties of $FM-S$ systems, where $FM$ is a
ferromagnetic metal \cite{Golubov}. The critical temperature of 
$FM-S-FM$ systems was studied recently in Ref. \cite{Tagirov}, 
where the (singlet)
superconducting film and ferromagnetic metals are assumed to be dirty, but
with (anti)parallel orientation of magnetization only.

The theoretical works above study singlet superconductors in contact with
ferromagnetic metals where the superconducting proximity effect dominates.
Here, we consider $FI-S-FI$ sandwiches where $FI$ stands for a ferromagnetic
insulator (semiconductor). In such systems conduction electrons penetrate the magnetic
layers on much smaller distances than in the case of metals and are totally
reflected at the $FI-S$ boundary. Additionally, the boundaries in $FI-S-FI$
systems are magnetically active and rotate spins of reflected electrons. 
The $FI-S-FI$
systems, with $FI$ a ferromagnetic insulator, might be of practical interest.
In contrast to ferromagnetic metals, where the proximity effect is
pronounced, this effect is drastically reduced in ferromagnetic insulators
(semiconductors) and the physics depends on fewer parameters. 
Note, the ferromagnetic semiconductors are already realized in systems like 
$EuO$, $EuS$ with the Curie temperature $T_{F}=66.8$ $K$ and $16.3$ $K$, 
respectively. In the magnetic semiconductor $EuSe$, which shows metamagnetic 
behavior, the antiferromagnetic and ferromagnetic order are realized at 
$T_{AF}=4.6$ $K$ and $T_{F}=2.8$ $K$.  

In this paper the critical temperature of the $FI-S-FI$ sandwich with 
singlet and triplet
superconductors is calculated for various orientations of the magnetization
in the $FI$ films. It turns out that the effect of magnetically
active boundaries on the superconducting critical temperature $T_{c}$
strongly depends on the orientation of magnetization. For the parallel 
orientation, for instance, $T_{c}$ is reduced much more than for 
the perpendicular one.
This property opens the possibility of
switches and logic operations based on $FI-S-FI$ systems, which is also briefly
discussed here.

\section{Theory of $FI-S-FI$ systems}

In the following we study systems consisting of a singlet or a triplet
superconductor and a ferromagnetic insulator as shown in $Fig.1$. 
\newline
\epsfysize=4.1in 
\hspace*{3.5cm} 
\vspace*{0cm} 
\epsffile{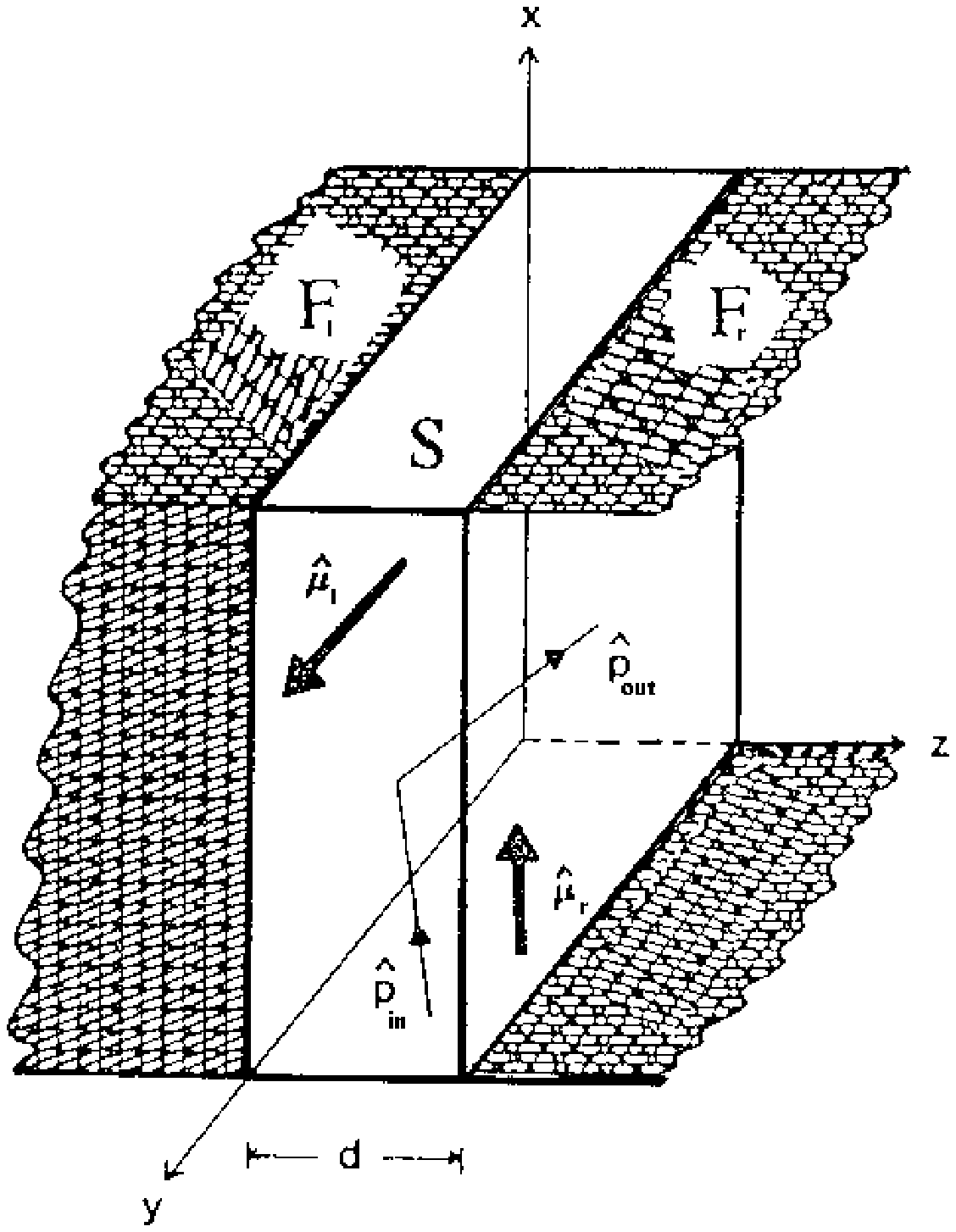} 
\newline
\mycap{{\bf FIGURE~1.} Thin film ($d\ll \xi _{0}$) geometry of a $FI-S-FI$ 
system. A superconducting film
(S) is sandwiched between two ferromagnetic insulators ($F_{l}$,$F_{r}$). 
The magnetization, {\mbox{\boldmath${\mu}$}$_{l,r}$}, is parallel to the 
$xy$-plane. The subscripts label the left(l) and the right(r) side of the 
sandwich. ${\bf \hat{p}}_{in}$ and ${\bf \hat{p}}_{out}$ refer to incoming 
and specular reflected (outgoing) trajectories, respectively.}

At the boundaries superconducting electrons penetrate the ferromagnetic
insulator only on short distances and are totally reflected. During this
short penetration time the electron spin is rotated by the exchange
field, ${\bf h}_{ex}$, of the ferromagnetic insulator. The $FI-S$ boundary can
be described by appropriate boundary conditions (see $Eq.(\ref{bound}$ %
)below) for the quasiparticle Green's function $\hat{g}({\bf \hat{p}}_{F},{\bf R}%
,\omega _{n})$, which were introduced in Ref. \cite{Millis} and applied 
to a $FI-S-I$ ($I$ is a nonmagnetic
insulator) sandwich in Ref. \cite{Toku}.

In this paper it is assumed that the exchange fields, ${\bf h}_{ex,l}$ and $%
{\bf h}_{ex,r}$, in the ferromagnetic insulator as well as the corresponding
magnetization, characterized by unit vectors {\boldmath{$\mu$}}$_{l}$ and {%
\boldmath{$\mu$}}$_{r}$({\bf \ }${\bf h}_{ex,l,r}\parallel $ 
{\boldmath{$\mu$}}$_{l,r}$), are parallel to the boundaries, but otherwise have
arbitrary orientation.

We describe the superconducting film by the quasiclassical equations of
Eilenberger \cite{Eilenberger} and Larkin-Ovchinnikov \cite{Larkin},
generalized to problems where spins of Cooper pairs are affected by magnetic
perturbations \cite{BuBuKuPa}. In the presence of magnetic perturbations
acting on quasiparticle spins, the quasiclassical Green's function is a $%
4\times 4$ matrix, $\hat{g}({\bf \hat{p}},{\bf R},\omega _{n})$, in the spin$%
\otimes$particle-hole product space 
\begin{equation}
\hat{g}({\bf \hat{p}},{\bf R},\omega _{n})=\left( 
\begin{array}{cc}
\tilde{g}({\bf \hat{p}},{\bf R},\omega _{n}) & \tilde{f}({\bf \hat{p}},{\bf R%
},\omega _{n}) \\ 
\tilde{f}(-{\bf \hat{p}},{\bf R},\omega _{n})^{\ast } & \tilde{g}^{tr}(-{\bf 
\hat{p}},{\bf R},-\omega _{n})
\end{array}
\right) ,  \label{gclass}
\end{equation}
where the normal and anomalous Green's functions, $\tilde{g}({\bf \hat{p}},
{\bf R},\omega _{n})$ and $\tilde{f}({\bf \hat{p}},{\bf R},\omega _{n})$, 
are $2\times 2$ matrices in spin space (see Ref. \cite{Serene}) 
\[
\tilde{g}=\tilde{g}_{s}+{\bf g}_{t}{\mbox{\boldmath${\sigma}$}} 
\]
\begin{equation}
\tilde{f}=(f_{s}+{\mbox{\boldmath${f}$}}_{t}{\mbox{\boldmath${\sigma}$}}
)i\sigma _{2}.  \label{gf}
\end{equation}
Here, the subscripts $s$ and $t$ correspond to singlet and triplet
superconductivity, respectively. $\sigma_{i}$ are Pauli matrices in spin space,
${\bf \hat{p}}={\bf p}_{F}/p_{F}$ and ${\bf p}_{F}$ is the Fermi momentum.
In the following, we assume the weak coupling limit for a clean
superconductor. The equation for $\hat{g}$ then reads 
\begin{equation}
\lbrack i\omega _{n}\hat{\tau}_{3}-\hat{\Delta}({\bf \hat{p}},{\bf R},\omega
_{n}),\text{ }\hat{g}({\bf 
\hat{p}},{\bf R},\omega _{n})]  \label{eloeq}
\end{equation}
\[
+\  i{\bf v}_{F} {\bf \nabla }_{{\bf R}}\hat{g}({\bf \hat{p}},{\bf R}%
,\omega _{n})=0, 
\]
where $\hat{\tau}_{i}=1 \otimes \tau_{i}$ and $\tau_{i}$, $i=1,2,3$, 
are Pauli matrices in particle-hole space. For later purposes we also define
$\hat{\sigma}_{i}=1\otimes \sigma_{i}$ for $i=1,2,3$.
The normalization condition for $\hat{g}$ is given by 
\begin{equation}
\hat{g}^{2}({\bf \hat{p}},{\bf R},\omega _{n})=-\hat{1},  \label{norm}
\end{equation}
while the order parameter $%
\hat{\Delta}({\bf \hat{p},R})$ is the solution of the self-consistency 
equation 
\begin{equation}
\hat{\Delta}({\bf \hat{p},R})=N(0)T\sum_{n}\int \frac{d\Omega ^{\prime }}{%
4\pi }V{\bf (\hat{p},\hat{p}}^{\prime }{\bf )}\hat{f}({\bf \hat{p}}^{\prime
},{\bf R},\omega _{n}).  \label{selfcons}
\end{equation}
Here, $V{\bf (\hat{p},\hat{p}}^{\prime }{\bf )}$ is the pairing potential,  $%
\hat{f}({\bf \hat{p}},{\bf R},\omega _{n})$ is the off-diagonal
part of $\hat{g}({\bf \hat{p}},{\bf R},\omega _{n})$ and $\Omega $
characterizes integration over the Fermi surface. The structure of the $%
4\times 4$ matrix $\hat{\Delta}$, which is different for singlet and triplet
pairing, will be given below.

To the equations above one needs to add boundary conditions for $\hat{g}(%
{\bf \hat{p}},{\bf R},\omega _{n})$ on the left and on the right $FI-S$
surface ${\bf R}_{s}^{l,r}$, respectively. These relate incoming, ${\bf \hat{%
p}}_{in}$, and outgoing, ${\bf \hat{p}}_{out}$, quasiparticles \cite{Toku} 
\begin{equation}
\hat{g}({\bf \hat{p}}_{in},{\bf R}_{s}^{l,r},\omega _{n})=\hat{S}_{l,r}\hat{g%
}({\bf \hat{p}}_{out},{\bf R}_{s}^{l,r},\omega _{n})\hat{S}_{l,r}^{-1}.
\label{bound}
\end{equation}
In the following we assume totally (specular) reflecting boundaries. $%
{\bf \hat{p}}_{in}$ and ${\bf \hat{p}}_{out}$ are then related by 
\begin{equation}
{\bf \hat{p}}_{out}={\bf \hat{p}}_{in}-2 {\bf {\hat{z}(\hat{p}_{in} \hat{z})}%
.}  \label{specular}
\end{equation}
The unit vector ${\bf \hat{z}}$ is normal to the $FI-S$ surface - see $Fig.1$.
The boundary scattering (rotation) matrix$\ \hat{S}\equiv \hat{S}%
({\bf \hat{p}}$,{\boldmath{$\mu$}}), which \
characterizes the magnetically active $FI-S$ surface, has the form\cite
{Millis},\cite{Toku}

\begin{equation}
\begin{array}{ccc}
\hat {S} & = & \left (
\begin{array}{cc}
\tilde{S}({\bf \hat{p}},{\mbox{\boldmath${\mu}$}}) & 0 \\ 
0 & \tilde{S}^{\ast }(-{\bf \hat{p}},-{\mbox{\boldmath${\mu}$}} )
\end{array}
\right ) \\ \\
& = & \left ( \begin{array}{cc} 
e^{-i\frac{\Theta}{2}{\mbox{\boldmath${\mu}$}}{\mbox{\boldmath${%
\sigma}$}}} & 0 \\ 
0 & e^{-i \frac{\Theta}{2} ({\mbox{\boldmath${\mu}$}} {%
\mbox{\boldmath${\sigma}$}})^{tr}}
\end{array}
\right).
\end{array} \label{smatrix}
\end{equation}

Here, $\hat{S}$ is the $4\times 4$ scattering matrix ($\tilde{S}$ is the $%
2\times 2$ spin scattering matrix) which describes rotation in spin space,
i.e. it rotates spins around the vector {\boldmath${\mu}$} by the
spin-mixing angle $\Theta $. The mixing angle depends on
physical quantities in the superconducting and the magnetic layer, like for
instance ${\bf \hat{p}}$, the exchange field, $h_{ex}$, and
the semiconducting gap, $E_{g}$. A model calculation for $\Theta ({\bf \hat{p%
}},h_{ex}, E_{g},p_{F})$ in a ferromagnetic semiconductor
is given in Ref. \cite{Toku}. For a ferromagnetic material with a large
semiconducting band gap one has $\Theta \sim (h_{ex}/E_{g})\ll1$. For our
purposes the mixing angle is considered a phenomenological parameter.

In the following, we study a $FI-S-FI$ system with a thin superconducting 
film, $%
d\ll \xi _{0}$, where $\xi _{0}$ is the superconducting correlation length.
In that case the solution of $Eq.(\ref{eloeq})$ is searched for in the form 
\begin{equation}
\hat{g}({\bf \hat{p}},{\bf R},\omega _{n})\approx \hat{g}_{0}({\bf \hat{p}}%
,\omega _{n})+(z-\frac{d}{2})\hat{g}_{1}({\bf \hat{p}},\omega _{n}),
\label{gsolu}
\end{equation}
with $\mid \hat{g}_{0}\mid \gg \mid d\hat{g}_{1}\mid $. \ By using the
boundary conditions given in $Eq.(\ref{bound})$ one can eliminate $\hat{g}%
_{1}$ in terms of $\hat{g}_{0}$, $\hat{S}_{l}$ and $\hat{S}_{r}$, which
leads to an equation for $\hat{g}_{0}$%
\[
\{[i\omega _{n}\hat{\tau}_{3}-\hat{\Delta},\hat{g}_{0}],\hat{S}%
_{l}\hat{S}_{r}\}+2i\frac{v_{F}\mid \hat{p}_{z}\mid }{d}[%
\hat{g}_{0},\hat{S}_{l}\hat{S}_{r}]+
\]
\begin{equation}
+\hat{S}_{l}[i\omega _{n}\hat{\tau}_{3}-\hat{\Delta},\hat{S}_{l}^{\dagger }\hat{g}_{0}\hat{S}_{l}+\hat{S}%
_{r}^{\dagger }\hat{g}_{0}\hat{S}_{r}]\hat{S}_{r}=0  \label{gnul}
\end{equation}
The brackets $\{..,..\}$ and $[..,..]$ mean
anticommutator and commutator, respectively. In the following, we solve $Eq.(%
\ref{gnul})$ near the critical temperature, $T_{c}$, for singlet as well as
triplet superconductors and for various orientations of {\boldmath{$\mu $}}$%
_{l}$ and {\boldmath{$\mu $}}$_{r}$.

\subsection{$T_{c}$ of $FI-S-FI$ systems with singlet superconductivity}

In the case of singlet superconductivity the $4\times 4$ matrix $\hat{\Delta}
$ is given by 
\begin{equation}
\hat{\Delta}=i\Delta _{s}\hat{\sigma}_{2}\hat{\tau}_{1}.  \label{deltas}
\end{equation}

Here, $\Delta_{s}$ is chosen real and in the following we 
omit the subscript $s$, i.e. $\Delta _{s}\equiv \Delta $.
Near $T_{c}$ the solution for $\hat{g}_{0}$ in $Eq.(\ref{gnul})$ is searched
for in the form 
\begin{equation}
\hat{g}_{0}=\hat{g}_{0}^{(0)}+\hat{f}_{0}^{(1)},  \label{gnulapprox}
\end{equation}
where $\hat{g}_{0}^{(0)}$ is independent of the order parameter, $\Delta $,
while $\hat{f}_{0}^{(1)}$ is linear in $\Delta $. From $Eq.(\ref{norm})$ it
follows 
\begin{equation}
\lbrack \hat{g}_{0}^{(0)}]^{2}\cong -\hat{1},  \label{gnulnorm}
\end{equation}
\begin{equation}
\{\hat{g}_{0}^{(0)},\hat{f}_{0}^{(1)}\}=0.  \label{gnulfnul}
\end{equation}

For a singlet superconductor the relative orientation of magnetization
is relevant and therefore the transition temperature only depends on the 
relative orientation of the exchange fields. Below,  $T_{c}$ is calculated 
for fields which are parallel, antiparallel or perpendicular to each other.
Particular orientations with respect to the $x$- and $y$-axis are chosen 
to perform the calculations. These choices are for convenience only and do not 
affect the final results. Furthermore, it is assumed that the mixing angles 
are identical, i.e. $\Theta _{l}=\Theta _{r}=\Theta $.

$1.$ $FI-S-FI$ sandwich with {\boldmath{$\mu$}}$_{l}=${\boldmath{$\mu$}}$_{r}=\bf
\hat{y}$

For parallel magnetization the rotation matrices at the left and the right 
boundary are the same, $\hat{S}_{l}=\hat{S}_{r}$, and $Eq.(%
\ref{gnul})$ then reduces to
\begin{equation}
\lbrack i\omega _{n}\hat{\tau}_{3}-\hat{\Delta}-\alpha \hat{\sigma}_{2}\hat{\tau}_{3},\hat{g}_{0}]=0,
\label{ssnul}
\end{equation}
with
\[
\alpha =\frac {v_{F}\mid \hat{p}_{z}\mid}{2d} \tan { \Theta}.
\]
$T_{c}$ is determined by
\begin{equation}
\ln t_{c}=-\sum_{n=0}^{\infty }\frac{1}{n+\frac{1}{2}}[1-(n+\frac{1}{2})%
\frac{t_{c}}{\rho}\arctan \frac{\rho}{t_{c}(n+%
\frac{1}{2})}].  \label{ssparallel}
\end{equation}
Here, $t_{c}=T_{c,\parallel }/T_{c0}$ and $\rho =\rho_{\mid 0} \tan {\Theta }%
/\tan (\Theta /2)$, where $\rho _{\mid 0}$ is defined by 
\begin{equation}
\rho _{\mid 0}=\frac{\xi _{0}}{2d}\tan \frac{\Theta }{2}.
\label{ronul}
\end{equation}
$T_{c0}$ is the critical temperature of the bulk and $\rho_{\mid 0}$ a 
parameter which describes pair-breaking for a sandwich with a single 
magnetically
active boundary, i.e. for the FI-S-I system($\mid 0$)
\cite{Toku}. If $\rho \ll 1$, one has 
$t_{c}=1-(7\zeta (3)/3)\rho^{2}=1-4(7\zeta (3)/3)\rho _{\mid 0}^{2}$, i.e. 
\begin{equation}
\frac{\delta T_{c,\parallel }}{T_{c0}} \equiv \frac {T_{c,\parallel}
-T_{c0}}{T_{c0}}=-\frac{28\zeta (3)}{3}\rho _{\mid 0}^{2}=
4 \frac {T_{c,\mid 0}-T_{c0}}{T_{c0}} 
\equiv 4 \frac{\delta T_{c,\mid 0}}{T_{c0}} \label{smallro}
\end{equation}
The functions $T_{c,\parallel }(\rho _{\mid 0})$ and 
$T_{c,\mid 0}(\rho _{\mid 0})$ are shown in $Fig.2$.

$2.$ $FI-S-FI$ with {\boldmath{$\mu$}}$_{l}=-${\boldmath{$\mu$}}$_{r}=\bf \hat{y}$

If the exchange fields are antiparallel the spin scattering matrices are 
related by $\hat{S}_{l} \hat{S}_{r}=\hat {1}$. As a result $Eq.(\ref{gnul})$ 
has the form 
\begin{equation}
\lbrack i\omega _{n}\hat{\tau}_{3}-a\hat{\Delta}-b\hat{\sigma}_{2}\hat{\Delta%
},\hat{g}_{0}]=0,  \label{x0s}
\end{equation}
where $a=\cos ^{2}(\Theta /2)$ and $b=i\sin(\Theta/2)$. The linearization
of $Eq.(\ref{x0s})$, as in $Eq.(\ref{gnulapprox})$, leads to 
\begin{equation}
T_{c,a}=T_{c0} e^{- \frac {\tan^{2}{\theta /2}}{\lambda }}.
\label{x-xs}
\end{equation}
Here, $\lambda $ is the superconducting coupling constant, i.e. $%
T_{c0}=1.13\omega _{c}e^{-1/\lambda }$. In the antiparallel case the effect
of ferromagnetic boundaries is pair-weakening, which means that $T_{c,a}$ goes to zero
when the pair-weakening parameter 
$\tan^{2}({\theta /2 })/\lambda \rightarrow \infty$. 
It is interesting to note that the
pair-weakening does not depend on the thickness
of the superconducting film in the limit $d\ll \xi _{0}$.
 However, for the more realistic case when $%
\Theta \ll 1$ the inequality $\mid \delta
T_{c,a} \mid \ll\mid\delta T_{c,\mid 0}\mid$ holds if $\lambda
\gg (d/\xi _{0})^{2}/3$. The condition on $\lambda $ is fulfilled for thin 
films of most
low-temperature superconductors. So, in the antiparallel case, $T_{c}$ 
is practically unchanged, i.e. $T_{c,a}\approx T_{c0}$. This property of the
$FI-S-FI$ system is in contrast to the antiparallel case of the $FM-S-FM$
systems where $T_{c,a}$ depends strongly on the thickness $d$ \cite{Tagirov}.
Note, the
same results also hold for the case when {\boldmath{$\mu $}}$_{l}$ and {%
\boldmath{$\mu $}}$_{r}$ are in opposite direction but along the $x$-axis.

$3.$ $FI-S-FI$ with {\boldmath{$\mu$}}$_{l}=\bf \hat{x}$ and 
{\boldmath{$\mu$}}$_{r}=\bf \hat{y}$

In the case of perpendicular exchange fields $Eq.(\ref{gnul})$ reads 
\begin{equation}
\lbrack i\omega _{n}\hat{\tau}_{3}-\hat{\Delta}-(\alpha _{1}\hat{\sigma}_{1}+\alpha _{2}\hat{\sigma}_{2}\hat{%
\tau}_{3}+\alpha _{3}\hat{\sigma}_{3}\hat{\tau}_{3}),\hat{g}_{0}]=0,
\label{gperp}
\end{equation}
where $\alpha _{i}=\alpha _{0}t_{i}$, $i=1-3$, $t_{1}=t_{2}=\tan \Theta /2$, 
$t_{3}=\tan ^{2}\Theta /2$, $\alpha _{0}=v_{F}\mid \hat{p}_{z}\mid /2d$.
Near $T_{c}$ the anomalous function $\hat{f}_{0}^{(1)}$ is searched for in
the form 
\[
\hat{f}_{0}^{(1)}=A\hat{\Delta}+B\hat{\tau}_{3}\hat{\Delta}+C\hat{\sigma}_{1}%
\hat{\Delta}+D\hat{\sigma}_{2}\hat{\Delta} 
\]
\begin{equation}
+F\hat{\sigma}_{1}\hat{\tau}_{3}\hat{\Delta}+G\hat{\sigma}_{2}\hat{\tau}_{3}%
\hat{\Delta}.  \label{fperp}
\end{equation}
Note, $A\equiv A(\omega _{n},\mid \hat{p}_{z}\mid )$ determines $T_{c}$.
Straightforward but cumbersome calculations give for $T_{c}$ 
\begin{equation}
\ln t_{c}=-c\sum_{n=0}^{\infty }\frac{1}{n+\frac{1}{2}}[1-(n+\frac{1}{2})%
\frac{t_{c}}{\rho}\arctan \frac{\rho}{t_{c}(n+\frac{1}{2}%
)}],  \label{tcperp}
\end{equation}
where $c=1/[1+(\tan ^{2}\Theta /2)/2]$, $\rho=
\sqrt{2}\rho _{\mid 0}/\sqrt{c}$. For $\rho\ll 1$ one obtains 
\begin{equation}
\delta T_{c,\perp }=2\delta T_{c,\mid 0}=\frac{\delta T_{c,\parallel }}{2}.
\label{dtcperp}
\end{equation}

In $Fig.2$ $T_{c}(\rho _{\mid 0})$ is shown for the cases when {\boldmath
{$\mu $}}$_{l}$ and {\boldmath{$\mu $}}$_{r}$ are parallel or perpendicular 
to each other and for the situation that only one boundary is magnetically 
active. 
\newline
\epsfysize=4.1in 
\hspace*{1cm} 
\vspace*{0cm} 
\epsffile{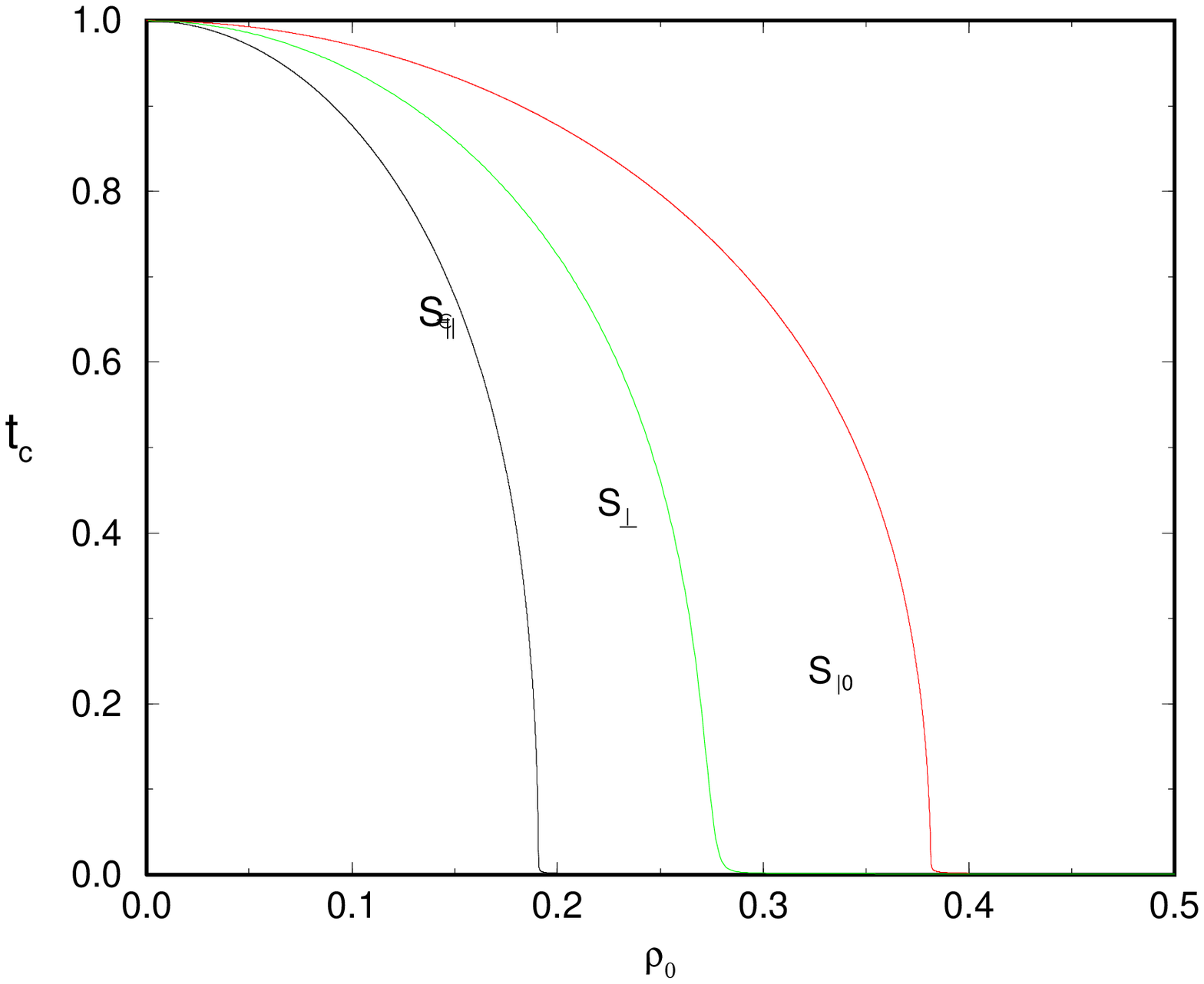} 
\newline
\mycap{{\bf FIGURE~2.} The critical temperature $t_{c}(\rho _{\mid 0})$ 
of a $FI-S-FI$ sandwich with a thin ($d\ll \xi _{0}$) singlet superconductor
and various orientations of the exchange fields ${\bf h}_{ex,l,r}$; $S_{\parallel} $ - parallel orientation; 
$S_{\perp} $ -
perpendicular orientation; $S_{|0}$ - the $FSI$ system. Note,
$\rho _{0}\equiv \rho _{\mid 0}$ given by $Eq.(\ref{ronul})$.}

\subsection{$T_{c}$ of $FI-S-FI$ systems with triplet superconductivity}

Let us consider $FI-S-FI$ systems with triplet superconductivity in which case
the $\hat{\Delta}$ matrix reads 
\begin{equation}
\hat{\Delta}=\left( 
\begin{array}{cc}
0 & ({\bf \Delta }_{t}\cdot {\mbox{\boldmath${\sigma}$}})i\sigma _{2} \\ 
i\sigma _{2}({\bf \Delta }_{t}^{\ast }\cdot {\mbox{\boldmath${\sigma}$}}) & 0
\end{array}
\right) .  \label{deltatrip}
\end{equation}
Between various possible triplet states we choose the pairing potential in
$Eq.(\ref{selfcons})$ which gives real $%
{\bf \Delta }_{t}={\bf \Delta }_{t}^{\ast }$,
with ${\bf \Delta }_{t}=\Delta ({\bf \hat{p}}){\bf \hat{x}}$ (Note, $%
\Delta (-{\bf \hat{p}})=-\Delta ({\bf \hat{p}})$.). This 
means that  ${\bf \Delta }_{t}$ is parallel to the
boundaries. The order parameter $\hat{\Delta}$ in this case is given by
\begin{equation}
\hat{\Delta}=-i\Delta ({\bf \hat{p}})\hat{\sigma}_{3}\hat{\tau}_{2}.
\label{deltatx}
\end{equation}
Note, if ${\bf \Delta }_{t}=\Delta ({\bf \hat{p}}){\bf \hat{y}}$ 
is realized the
physics is similar but with appropriately chosen orientation of
magnetization at the boundaries. Since the physics of
the problem does not depend on $\Delta ({\bf \hat{p}})$, we take $\Delta 
({\bf \hat{p}})=\hat{p}_{x} \Delta$, 
where $\Delta$ is constant, to simplify the calculation.

For triplet pairing the absolute orientation of ${\bf h}_{ex,l}$($\parallel${%
\boldmath{$\mu$}}$_{l}$) and ${\bf h}_{ex,r}$($\parallel${\boldmath
{$\mu$}}$_{r}$) plays a very important role and different results are
obtained when ${\bf h}_{ex,l,r}$ are orientated along the $x$- or $y$-axis.
If, for instance, both fields are perpendicular to the order parameter 
${\bf \Delta }_{t}$ the
transition temperature is the same as in a bulk superconductor, while in other cases it is not - see below. This property opens a new possibility for testing
the structure of the order parameter in triplet superconductors - see discussions below. 
As above, the mixing angles are assumed equal($\Theta _{l}=\Theta _{r}=\Theta$).

$1.$ $T_{c}$ of a $FI-S-FI$ sandwich with {\boldmath{$\mu$}}$_{l}=${\boldmath{$%
\mu$}}$_{r}=${\boldmath{$\hat{x}$}}

If the exchange fields are both parallel to the $x$-axis $\hat{g}_{0}$
fulfills $Eq.(\ref{gperp})$. After linearization one obtains an equation for 
$T_{c}$%
\begin{equation}
\ln t_{c}=-\frac{3}{4}\sum_{n=0}^{\infty }\frac{1}{n+\frac{1}{2}}[I(a_{n})-%
\frac{2}{3}]  \label{ttpar}
\end{equation}
with 
\begin{equation}
I(a_{n})=2(1+a_{n}^{2})(1-a_{n}\arctan \frac{1}{a_{n}}).  \label{I}
\end{equation}
$a_{n}=(n+1/2)t_{c}/\rho$ and $\rho=(\tan \Theta
/\tan \Theta /2)\rho _{\mid 0}$ with $\rho _{\mid 0}$ given by $Eq.
(\ref{ronul})$. The
function $T_{c,\parallel }(\rho _{\mid 0})$ is shown in $Fig.3$. However, in 
the case when $\rho \ll 1$ one has $\rho =2\rho _{\mid 0}$ and $Eq.
(\ref{ttpar})$ can be solved analytically. The result for $\delta T_
{c,\parallel }(\equiv T_{c,\parallel}-T_{c0})$ reads

\begin{equation}
\frac{\delta T_{c,\parallel }}{T_{c0}}=-\frac{7\zeta (3)}{5}\rho^{2}=
-\frac{28\zeta (3)}{5}\rho _{\mid 0}^{2}  \label{ttsmall}
\end{equation}
Assuming that $T_{c0}$ for singlet and triplet pairing is the same, this
result means that ferromagnetic boundaries are less detrimental for triplet
than for singlet pairing - compare $Eq.(\ref{smallro})$ and $Eq.(\ref
{ttsmall})$. Note, in a $FI-S-I$ system with triplet superconductivity and with
a magnetization which is parallel to the order parameter $Eq.(\ref{ttpar})$
holds if $\rho$ is replaced by $\rho _{\mid 0}$.

$2.$ $T_{c}$ of a $FI-S-FI$ sandwich with {\boldmath{$\mu$}}$_{l}=-${\boldmath{$%
\mu$}}$_{r}=$ {\boldmath{$\hat{x}$}}

In the antiparallel case the ferromagnetic boundaries are pair-weakening
and $T_{c}$ is given by the same expression, $Eq.(\ref{x-xs})$, as for 
singlet pairing. The pair-weakening parameter does again not depend on the thickness of the superconducting film  
in the limit $d\ll \xi _{0}$.

$3.$ $T_{c}$ of a $FI-S-FI$ sandwich with {\boldmath{$\mu$}}$_{l}=$ {\boldmath{$%
\hat{x}$}} and {\boldmath{$\mu$}}$_{r}=$ {\boldmath{$\hat{y}$}}

In the perpendicular ($\perp $) geometry $Eq.(\ref{gperp})$ holds with $\hat{%
\Delta}$ from $Eq.(\ref{deltatrip})$. $\hat{f}_{0}^{(1)}$ is searched for in
the form of $Eq.(\ref{fperp})$. $T_{c}$ is given by

\begin{equation}
\ln t_{c}=-\frac{3}{2}\sum_{n=0}^{\infty }\frac{1}{n+\frac{1}{2}}%
\int_{0}^{1}(1-x^{2})\frac{b_{n}(x)}{1+b_{n}(x)}dx,  \label{ttperp}
\end{equation}
\begin{equation}
b_{n}(x)=x^{2}\frac{(n+\frac{1}{2})^{2}y^{2}+2x^{2}}{(n+\frac{1}{2}%
)^{4}y^{4}+3(n+\frac{1}{2})^{2}y^{2}x^{2}+2x^{4}},  \label{bn}
\end{equation}
with $y=t_{c}/\rho _{\mid 0}$. For $\rho _{\mid 0}\rightarrow 0$ one obtains 
\[
\frac{\delta T_{c,\perp }}{T_{c0}}=-\frac{7\zeta (3)}{5}\rho _{\mid 0}^{2}, 
\]
i.e. $\delta T_{c,\parallel }=4\delta T_{c,\perp }$. The calculations above 
show that the ferromagnetic boundaries are much less detrimental for the
perpendicular geometry than for the parallel one. This can also be seen in $%
Fig.3$ where it is apparent that $T_{c,\parallel }$ vanishes at smaller $%
\rho _{\mid 0}$ than $T_{c,\perp }$. It is interesting to note that in the
perpendicular case there is a reentrant behavior of triplet
superconductivity, i.e. superconductivity disappears in some interval $%
T_{c,\perp }^{(2)}<T<T_{c,\perp }^{(1)}$ while it reappears at $T<T_{c,\perp
}^{(3)}$, where $T_{c,\perp }^{(3)}<T_{c,\perp }^{(2)}<T_{c,\perp }^{(1)}$.
This result might mean that instead of the assumed second-order phase
transition a first-order phase transition takes place in some region of the $%
(T_{c,\perp },\rho _{\mid 0})$ phase diagram. 
\newline
\epsfysize=4.1in 
\hspace*{1cm} 
\vspace*{0cm} 
\epsffile{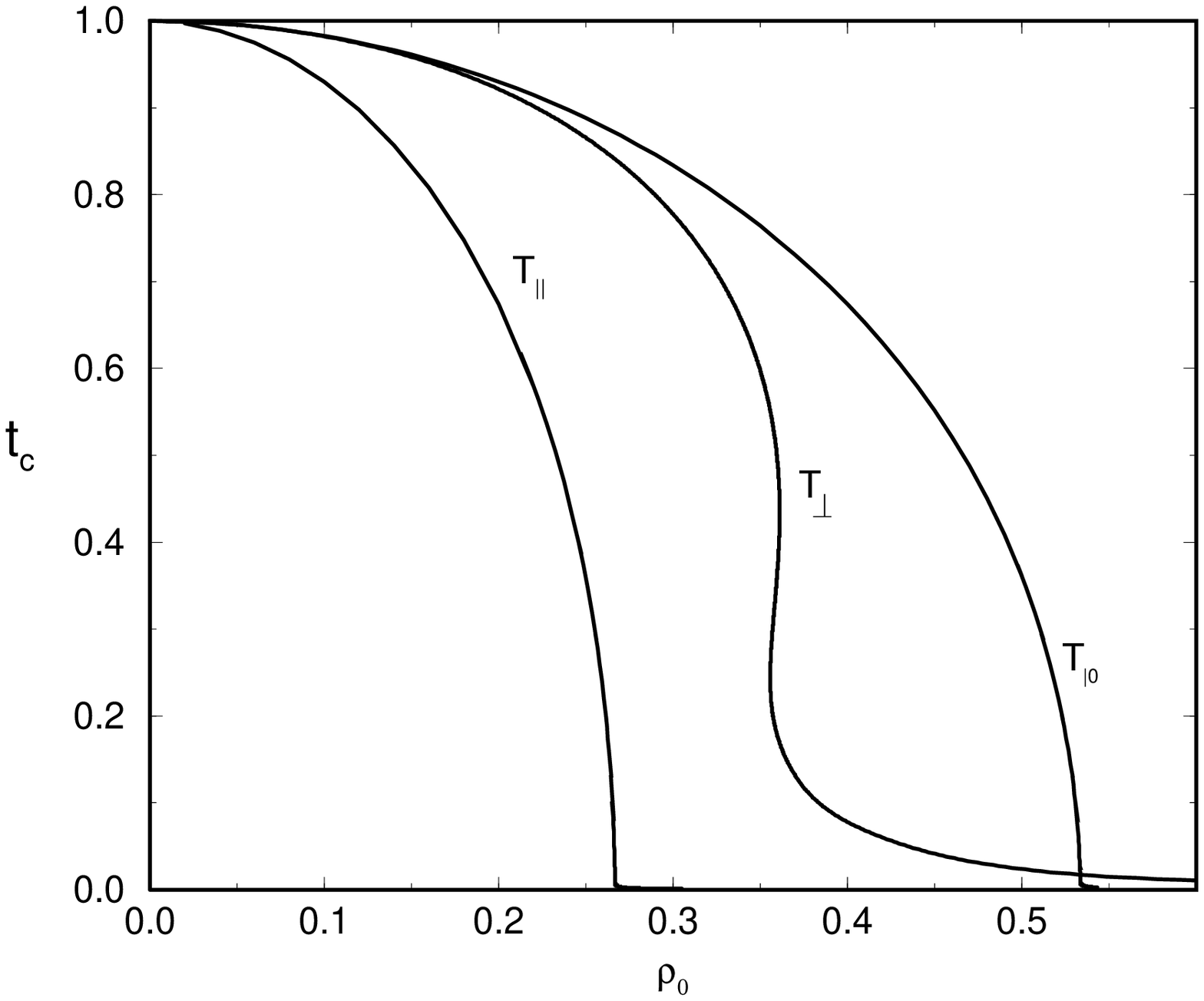} 
\newline
\mycap{{\bf FIGURE~3.} The critical temperature $t_{c}(\rho _{0})$ 
of a $FI-S-FI$ sandwich for a thin($d\ll \xi _{0}$)triplet superconductor 
and various orientations of the exchange fields, ${\bf h}_{ex,l,r}$; $T_{\parallel} $ - parallel orientation; 
$T_{\perp} $ -
perpendicular orientation; $T_{|0}$ - the $FSI$ system with magnetization 
along the $x$-axis.  Note,
$\rho _{0}\equiv \rho _{\mid 0}$ given by $Eq.(\ref{ronul})$.}

\section{Discussion}

We have shown that the critical temperature of a ferromagnetic semiconductor -
superconductor - ferromagnetic semiconductor sandwich depends on
the orientation of the exchange fields, ${\bf h}_{ex,l,r}$. In case of singlet
superconductivity $T_{c}$ depends on their relative orientation only, while 
for triplet superconductivity it depends on their absolute orientation.
Significant
depression of $T_{c}$ for singlet superconductors starts for a pair-breaking
parameter $2\rho _{\mid 0}\geq 0.3$ if the exchange fields
are parallel to each other. For such values of $\rho _{\mid 0 }$ there is
significant anisotropy of $T_{c}$ in this system, i.e. $T_{c,\parallel
}<T_{c,\perp }<T_{c,\mid 0}<T_{c,a}\simeq T_{c0}$ as shown in $Fig.2$.
Hence, the superconductor can be switched from its normal to its
superconducting state or vice versa by changing the relative orientation of
the magnetization in the magnetic layers. If, for instance, the exchange
fields of a $FI-S-FI$ sandwich (with a singlet superconductor) are perpendicular
to each other at a given temperature $T$, which
fulfills $T_{c,\parallel }<T<T_{{c,\perp }}$, the superconductor is in its
superconducting state. Rotation of one of the fields by ninety degrees to the
parallel configuration then
switches the superconductor to its normal state. This means that in the
perpendicular configuration the state ''1'' is realized, while in the
parallel one the state ''0'' is realized. Depending on the magnetic
properties of the ferromagnetic insulator (like for instance magnetic
anisotropy) it may happen that energy losses of such a switch are minimized
if it operates between the perpendicular and the parallel configuration. By
combining many such switches various logic circuits can be realized what
will be analyzed elsewhere.

The reorientation of magnetization in thin ferromagnetic films is already
realized experimentally \cite{Zabel3}. This
result opens the possibility of $FI-S-FI$ switches.

There is another possibility of a realization of logic circuits. If one
keeps, for instance, the temperature fixed, $T<T_{c,\parallel }<T_{c,\perp
}<T_{c,a}$, in a cryotron-like device, which consists of several $FI-S-FI$
switches with fixed orientation of the exchange fields, ${\bf h}_{ex,l,r}$, one
can then reach that some switches pass to normal state while others stay in
the superconducting state by changing the current in the control device.
Various designs based on these switches are possible (imaginable), also by
combining these two possibilities .

Similar applications are possible with $FI-S-FI$ switches based on triplet
superconductivity as well as with combined singlet and triplet
superconducting $FI-S-FI$ switches.

Note, triplet superconductivity is probably realized in the layered
perovskite superconductor $Sr_{2}RuO_{4}$ \cite{Maeno} with $T_{c}\cong 1.5$ 
$K$, while the structure of the order parameter is still unknown. However,
there is some evidence that the order parameter belongs to the $E_{u}$
irreducible representation of the $D_{4h}$ group, i.e. ${\bf \Delta }_{t}(%
{\bf \hat{p}})=\Delta (\hat{p}_{a}\pm i\hat{p}_{b}){\bf \hat{c}}$, where the
unit vectors ${\bf \hat{a}},{\bf \hat{b}}$ lie in the $RuO_{2}$ plane and $%
{\bf \hat{c}}$ is perpendicular to it \cite{Ishida}. 
In the analysis of $FI-S-FI$ systems with triplet superconductivity it is 
shown that $T_{c}$ strongly depends on the absolute orientation of 
magnetization with respect to the order parameter 
${\bf \Delta }_{t}$. According to this result we expect that similar behavior
will be realized in $Sr_{2}RuO_{4}$ with the order parameter proposed above 
\cite{Ishida}.
The possibility of
resolving its structure by using $FI-S-FI$ systems will be analyzed elsewhere.

In conclusion, the $FI-S-FI$ switches and logic circuits would be advantageous
compared to $FM-S-FM$ devices, because in the former fewer physical 
parameters need to be controlled 
than in the latter. The $FI-S-FI$ systems can be used in
spin-dependent superconducting spectroscopy especially
in resolving the 
structure of the order parameter in triplet and unconventional
superconductors like, for instance, the newly discovered $Sr_{2}RuO_{4}$.

{\bf Acknowledgments}

It is a pleasure for us to thank Hartmut Zabel for very
useful discussions on $FI-S-FI$ systems, after which the idea for this paper arose.
We thank Dierk Rainer for useful discussions, comments and reading
of the manuscript. 
M. L. K. gratefully acknowledges the support of the
Deutsche Forschungsgemeinschaft through the Forschergruppe
''Transportph\"{a}nomene in Supraleitern und Suprafluiden''.

\end{document}